\documentclass[twocolumn,aps,prc,superscriptaddress]{revtex4}

\usepackage{amssymb}
\usepackage{amsmath}
\usepackage{graphicx}
\usepackage[normalem]{ulem}
\usepackage{multirow}
\usepackage{appendix}
\usepackage{CJK}
\usepackage[usenames]{color}
\usepackage{bm}
\mathchardef\mhyphen="2D

\makeatletter 
\@addtoreset{equation}{section}
\makeatother  
\usepackage{hyperref}
\makeatletter
\def\UrlAlphabet{%
      \do\a\do\b\do\c\do\d\do\e\do\f\do\g\do\h\do\i\do\j%
      \do\k\do\l\do\m\do\n\do\o\do\p\do\q\do\r\do\s\do\t%
      \do\u\do\v\do\w\do\x\do\y\do\z\do\A\do\B\do\C\do\D%
      \do\E\do\F\do\G\do\H\do\I\do\J\do\K\do\L\do\M\do\N%
      \do\O\do\P\do\Q\do\R\do\S\do\T\do\U\do\V\do\W\do\X%
      \do\Y\do\Z}
\def\UrlDigits{\do\1\do\2\do\3\do\4\do\5\do\6\do\7\do\8\do\9\do\0}
\g@addto@macro{\UrlBreaks}{\UrlOrds}
\g@addto@macro{\UrlBreaks}{\UrlAlphabet}
\g@addto@macro{\UrlBreaks}{\UrlDigits}
\makeatother

\begin{document}
\begin{CJK*}{GB}{gbsn}
\title{	Searching for $\rm{\overline{{}^4Li}}$ via the momentum correlation function of $\rm{\overline{p}}$-$\rm{\overline{{}^3He}}$}
\author{Bao-Shan Xi(Û­±¦É½)}
\affiliation{Shanghai Institute of Applied Physics, Chinese Academy of Sciences, Shanghai 201800, China}
\affiliation{University of Chinese Academy of Sciences, Beijing 100049, China}
\affiliation{ShanghaiTech University, Shanghai 201800, China}
\author{Zheng-Qiao Zhang(ÕÅÕýÇÅ)\footnote{Email: qiao@rcf.rhic.bnl.gov}}
\affiliation{Shanghai Institute of Applied Physics, Chinese Academy of Sciences, Shanghai 201800, China}
\author{Song Zhang(ÕÅËÉ)}
\affiliation{Key Laboratory of Nuclear Physics and Ion-beam Application (MOE), Institute of Modern Physics, Fudan University, Shanghai 200433, China}
\author{Yu-Gang Ma(ÂíÓà¸Õ)\footnote{Email: mayugang@fudan.edu.cn}}
\affiliation{Key Laboratory of Nuclear Physics and Ion-beam Application (MOE), Institute of Modern Physics, Fudan University, Shanghai 200433, China}
\affiliation{Shanghai Institute of Applied Physics, Chinese Academy of Sciences, Shanghai 201800, China}

\date{\today}

\begin{abstract}
The heaviest observed anti-nucleus to date is $\rm{\overline{{}^4He}}$ which was dectected at the STAR experiment at the Relativistic Heavy Ion Collider. From previous scattering experiment,  we know that the $\rm{{}^4Li}$ has a very short lifetime about $1.197\times10^{-22}s$, and can decay into proton and $\rm{{}^3{H}e}$. In experiment, the correlation function of  $\rm{\overline{p}}$-$\rm{\overline{{}^3He}}$ provides us a method to observe $\rm{\overline{{}^4Li}}$. In this paper, we use the blast-wave model and  Lednick\'y-Lyuboshitz analytical model to obtain a prediction of the correlation function of  $\rm{\overline{p}}$-$\rm{\overline{{}^3He}}$ with/without  $\rm{\overline{{}^4Li}}$ decay in Au + Au collisions at $\sqrt{S_{NN}}$ = 200 GeV. The magnitude of event number needed to detect $\rm{\overline{{}^4Li}}$ experimentally is estimated from the error of the correlation function. The correlation function with $\rm{\overline{{}^4Li}}$ decay is found to exhibit a peak at $k^* \approx$ 0.073 GeV/$c$. The results offer a reference for the experimental search for $\rm{\overline{{}^4Li}}$ in relativistic heavy ion collisions.
\end{abstract}

\maketitle

\section{Introduction}

Nucleus-nucleus collisions from a few GeV to the Large Hadron Collider (LHC) energy regime not only provides a hot-dense environment for understanding the properties of QCD matter \cite{NPA1,NPA2,NPA3,NPA4,PBM,LuoNST,SCI2,WangFQ} but also  produce abundant light nuclei, strange baryons and their corresponding anti-particles, and even hypernuclei or anti-hypernuclei \cite{STAR-Science,naturexue,STAR1,E864,ALICE-NP,STAR-NP,LiuP,Lea,sinap,Xue,Neha,Sun1,Sun2,Zhu,Cho}. Such collisions provide us an ideal venue to study the production of light nuclei and their anti-partners. 
Usually, detections of such (anti-)light nuclei and strange baryons  were performed by the invariant mass analysis or direct identification with specific energy loss of ions in tracking detectors methods. For instance,  the STAR Collaboration reported the observation of first anti-hypernucleus, namely ($\rm{\overline{{}^3_\Lambda H}}$)~\cite{STAR-Science} by the  invariant mass reconstruction   as well as  the  antimatter partner of  $\rm{^4He}$,  namely $\rm{\overline{{}^4He}}$ by  identification using specific energy loss of ions in tracking detectors~\cite{naturexue}, which is the heaviest anti-particle observed so far. The production yield of the next stable antimatter nucleus is $\rm{\overline{{}^6Li}}$, which has about eight orders of magnitude yield less than that of $\rm{\overline{{}^4He}}$, therefore it is almost not feasible   to detect $\rm{\overline{{}^6Li}}$ in current experiments~\cite{sinap}. 

However, for $\rm{\overline{{}^4Li}}$, which has almost the same mass as $\rm{\overline{{}^4He}}$, its yield is about 4 times larger than $\rm{\overline{{}^4He}}$ according to the thermal model which can offer a good estimate of particle yields in Au + Au collisions at $\sqrt{S_{NN}}$ = 200 GeV~\cite{PBM,thermal,thermal2}. Comparing to $\rm{\overline{{}^4He}}$,  $\rm{\overline{{}^4Li}}$ is unstable and has a very short lifetime. According to the symmetric property of matter and antimatter, from the lifetime of $\rm{^4Li}$, we can get that the lifetime of $\rm{\overline{{}^4Li}}$ is $1.197\times10^{-22}s$~\cite{lifeTimeLi4},  and it decays into $\rm{\overline{{}^3He}}$ and $\rm{\overline{p}}$.

Beside the method of invariant mass reconstruction, the $\rm{\overline{{}^4Li}}$ and $\rm{{}^4Li}$ yields can be also deduced from the $\rm{\overline{p}}$-$\rm{\overline{{}^3He}}$ and $\rm{p}$-$\rm{{}^3\!He}$ correlation functions, together with the information on their strong interaction forces~\cite{c71,extra,lamlam,fin,Neha2,inter}. The STAR Collaboration has already measured the momentum correlation function of two antiprotons and two protons to extract their interaction parameters and confirmed equal strong interaction in matter and antimatter~\cite{nature}. In present paper, we simulate the  $\rm{\overline{p}}$-$\rm{\overline{{}^3He}}$ and $\rm{p}$-$\rm{{}^3\!He}$ correlation functions with and without $\rm{\overline{{}^4Li}}$ and $\rm{{}^4Li}$ decays in Au + Au collisions at $\sqrt{S_{NN}}$ = 200 GeV and estimate the statistics required to observe $\rm{\overline{{}^4Li}}$ and $\rm{{}^4Li}$. Note that the measurement of the $\rm{{}^4Li}$ yield through the $\rm{p}$-$\rm{{}^3\!He}$ correlation function was considered in Refs. ~\cite{hbt87,li4h4} and the measurement of the $\rm{{}^4Li}/\rm{{}^4He}$ ratio in central and peripheral collisions at RHIC or LHC was suggested to discriminate between thermal and coalescence models of light nuclei production ~\cite{li4h4}.

 It is well known that particles produced by resonance decay will affect the correlation function of directly emitted particles~\cite{inter,inf}. By measuring the correlation function of the two particles from the resonance decay, the parent particle before the resonance decay could be found~\cite{hbt87}.
 On the other hand, searching for heavier antiparticles is always a  very interesting and important topic in both cosmic rays and heavy ion collisions since it helps to understand the  matter-antimatter asymmetry \cite{sinap}. In thermal model where the collision system can be considered as a fireball at an extremely high temperature, the production of light (anti)nucleus can be described by the Boltzmann factor $e^{-|B|m_p/T}$ where $|B|$ is the baryon number \cite{PBM,Neha,Xue}.

 In addition to the real experimental measurement, it is also useful to derive the correlation function and give a guidance for experiments by simulating the process of high energy heavy ion collisions with various models~\cite{fourier,inf,Neha,Xue,ther,Jin1,WangHM,Lao,Jin2,WangH}. In this paper the fireball formed after collisions of two nuclei are simulated through a blast-wave model~\cite{dragon},  which can  generate events through a Monte Carlo simulation and can deal with the resonance decay of emitted particles.
    
    In Lednick\'y-Lyuboshitz model, the weight due to the final state interactions (FSIs) of each pair from the phase space is calculated as the square of the properly symmetrized wavefunction averaged over the total pair spin and the distribution of relative distances of particle emission points in the pair rest frame ~\cite{LL,note,ll05}.  For a pair of particles composed of two particles, the calculation process of the correlation function may vary with the variety of particles~\cite{int,zero,deu}.
The momentum information of $\rm{\overline{p}}$ and $\rm{\overline{{}^3He}}$ comes from the blast-wave model, and the position information comes from the assumed Gaussian source. The radius of the Gaussian source depends on the centrality of collisions.
    The input parameters in the model for  $\rm{\overline{p}}$-$\rm{\overline{{}^3He}}$ correlation function come from previous $\rm{p}$-$\rm{{}^3\!He}$ scattering experiment ~\cite{sca,zero}. Thus we can compare the correlation functions from phase space with and without $\rm{\overline{{}^4Li}}$ emission.

	The rest of paper is organized as follows. Section \ref{sec:THEORY} briefly reviews the definition of correlation function and the method of obtaining correlation function in experiments and theory. Here the blast-wave model is used to generate the phase space of the fireball, and then Lednick\'y-Lyuboshitz model is applied to calculate the correlation function. In Section \ref{sec:RESULTS} the correlation function of $\rm{\overline{p}}$-$\rm{\overline{{}^3He}}$ is given and the results are discussed. A summary is given in Sec. \ref{sec:SUMMARY}.

    \section{Theoretical framework}
    \label{sec:THEORY}

    \subsection{Experimental correlation function }
    \label{subsec:cf}

    Experimentally the correlation function can be constructed by the ratio of the relative momentum distributions of correlated and uncorrelated particles, and it is influenced by quantum statistical effect and the final state interaction of particles. This method is widely used to study the space-time properties of emission source at the fermi scale. The two-particle correlation function in experiment can be obtained from the following formula
 \begin{equation}
    C(k^{*}) = \frac{A(k^{*})}{B(k^{*})}. 
    \end{equation}
Here $k^{*} = |\boldsymbol{k}^{*} |$ is the relative momentum of one of the particles in the pair rest frame ~\cite{Neha2,p_lambda}. 
	$A(k^{*})$ is the  $k^{*}$ distribution for correlated pairs from the same event, and $B(k^{*})$ is the $k^{*}$ distribution for uncorrelated pairs from two different events. Correlation function is sensitive to the size of the emission source and interaction between particles but not sensitive to the momentum distribution of a single particle and the detection efficiency of the detector \cite{Wei,Wang,WangTT2}.

	    \subsection{Lednick\'y-Lyuboshitz model}
	    
	    Correlation function is computed using the Lednick\'y-Lyuboshitz model.
	    Firstly, the $s$-wave scattering amplitude is obtained by 
	    \begin{align}
	    f^{S} \left( k ^ { * } \right) = \left[ \frac { 1 } { f ^{S}_ { 0 } } + \frac { 1 } { 2 } d ^{S}_ { 0 } k ^ { * 2 } - \frac { 2 } { a _ { c } } h \left( k ^ { * } a _ { c } \right) - i k ^ { * } A _ { c } ( \eta ) \right] ^ { - 1 },
	    \end{align}
	    where $f^{S}_{0}$ is the scattering length and $d^{S}_{0}$ is the effective range, which are two important parameters for describing strong interaction. The superscript $S$ is the total spin. $S$ = 0 and 1 denotes singlet and triplet, respectively. $A_{c}( \eta) = 2 \pi\eta [\exp(2\pi\eta)-1]^{-1}$ is the Coulomb penetration factor where $\eta = (k^{*}a_{c})^{-1}$ and $a_{c}=19.2$ fm is the Bohr radius for $\rm{\overline{p}}$ and $\rm{\overline{{}^3He}}$. And 
	    \begin{equation}
	    h ( x ) = \frac { 1 } { x ^ { 2 } } \sum _ { n = 1 } ^ { \infty } \frac { 1 } { n \left( n ^ { 2 } + x ^ { - 2 } \right) } - C + \ln | x |,
	    \end{equation}
	    where $C = 0.5772$ is the Euler constant.

	    In the $\rm{\overline{p}}$-$\rm{\overline{{}^3He}}$ pairs, the values of the parameters characterizing the strong interaction are set to $f_{0}^{(0)}$ = -11.1 fm and $d_{0}^{(0)}$ = 1.85 fm for the singlet state, $f_{0}^{(1)}$ = -9.05 fm and  $d_{0}^{(1)}$ = 1.68 fm for the triplet state~\cite{f0d0,newf0}.
	    
	    Next, according to approximation of the outer solution of the scattering problem~\cite{19,35}, the equal-time reduced Bethe-Salpeter amplitude is calculated as
	    \begin{align}
	    \notag \psi _ { \boldsymbol{-k} ^ { * } }^ { S(+) } \left( \boldsymbol{r} ^ { * } \right) = e ^ { i \delta _ { c } } \sqrt { A _ { c } ( \eta ) } \left[\right. &e ^ { - i \boldsymbol{k} ^ { * }\cdot\boldsymbol{r} ^ { * } } F ( - i \eta , 1 , i \xi ) \\ +& f _ { c } \left( k ^ { * } \right) \frac { \widetilde { G } ( \rho , \eta ) } { r ^ { * } } \left.\right]; \\
	    \notag \psi _ { \boldsymbol{k} ^ { * } }^ { S } \left( \boldsymbol{r} ^ { * } \right) = e ^ { i \delta _ { c } } \sqrt { A _ { c } ( \eta ) } \left[\right. &e ^ { i \boldsymbol{k} ^ { * }\cdot\boldsymbol{r} ^ { * } } F ( - i \eta , 1 , i(\rho-\boldsymbol{k} ^ { * }\cdot\boldsymbol{r} ^ { * }) ) \\ +& f _ { c } \left( k ^ { * } \right) \frac { \widetilde { G } ( \rho , \eta ) } { r ^ { * } } \left.\right].	    
	    \end{align}
	
	    Here $F$ is confluent hypergeometric function, $\rho=k ^ { * } r^ { * }$, $\xi=\boldsymbol{k} ^ { * }\cdot\boldsymbol{r} ^ { * }+\rho$. And
	    \begin{equation}
	    \widetilde { G } ( \rho , \eta ) = \sqrt { A _ { c } \left( \eta  \right) } \left( G _ { 0 } ( \rho , \eta ) + i F _ { 0 } ( \rho , \eta ) \right),
	    \end{equation}
	    where the $F_{0}$ is regular $s$-wave Coulomb function and the $G_{0}$ is singlet $s$-wave Coulomb function.
	    
	    With these terms, the weight of pair with $r^{*}$ and $k^{*}$ can be obtained as
	    \begin{align}
	    w \left( \boldsymbol{k} ^ { * } , \boldsymbol{r} ^ { * } \right) = \sum_{S}R _{S}\langle\left| \psi _ { - \boldsymbol{k} ^ { * } } ^ { S(+)  } \left( \boldsymbol{r} ^ { * } \right) \right| ^ { 2 }\rangle_{S},
	    \label{eq:equation133}
	    \end{align}
where we assume that particles are produced unpolarized, here $R_{0}$ is $\frac{1}{4}$ and $R_{1}$ is $\frac{3}{4}$ for the pairs
	    in the singlet state and the triplet state, respectively.
	    
	    At last, theoretical correlation function can be obtained by
	    \begin{align}
	    CF \left(k^{*} \right)=\frac{\sum_{pairs}\delta \left(k_{pair}^{*}-k^{*} \right) w \left(\boldsymbol{k}^{*},\boldsymbol{r}^{*}\right)}{\sum_{pairs} \delta \left(k_{pair}^{*}-k^{*} \right) }.
	    \label{eq:equation13}
	    \end{align}

    \subsection{Generation of phase space}
    \label{subsec:dragon}
    In blast-wave models, the phase-space information of emitted particles from the fragmented fireball can be obtained for Au+Au collisions at $\sqrt{S_{NN}}$ = 200 GeV including the ground state of $\rm{\overline{{}^4Li}}$~\cite{nuclear_data}. In this model, relative coordinates and polar coordinates are used to describe the position of particles. The phase-space distribution of hadrons emitted from the expanding fireball can be expressed as a Wigner function:
    \begin{equation}
    \begin{aligned} S(x, p) d^{4} x & = \frac{2 s+1}{(2 \pi)^{3}} m_{t} \cosh (y-\eta) \exp \left(-\frac{p^{\mu} u_{\mu}}{T_{k}}\right) \\ & \times \Theta(1-\tilde{r}(r, \phi)) H(\eta) \delta\left(\tau-\tau_{0}\right) d \tau \tau d \eta r d r d \phi \end{aligned}, 
    \end{equation}
    where the distribution $H(\eta)$ is related to the scale of fireball in space-time rapidity, $T_{k}$ is the kinetic
freeze-out temperature. $s$, $y$, and $m_t$ are the spin, rapidity, and transverse mass of the hadron, respectively, and $p^\mu$ is the four-component momentum. $\tau_{0}$ is the Bjorken lifetime, and 10.5 or 8 fm/$c$ is used for central or peripheral collisions, respectively. Equation (2) is formulated in a Lorentz covariant
way, $r$ and $\phi$ are the polar coordinates,  $\eta$ and $\tau$ are the pseudo-rapidity and the proper time, respectively.
    
   In radial direction, emission points are distributed uniformly
    \begin{equation}
    \tilde{r} = \sqrt{\frac{\left(x^{1}\right)^{2}}{R_{x}^{2}}+\frac{\left(x^{2}\right)^{2}}{R_{y}^{2}}}<1, 
	\end{equation}
	with ($x^1$,$x^2$) standing for the coordinates in the transverse
plane and $R_{x,y}$ being the average transverse radius, i.e. $R_{x} = a R$ and $\quad R_{y} = \frac{R}{a}$ with $R$ is the average transverse radius of an ellipsoid fireball and $a$ is spatial deformation parameter and here we set it to 1~\cite{dragon}. 

 The radial flow is
\begin{equation}
\left\langle\beta_{T}\right\rangle=\int \operatorname{arctanh}\left(\rho_{0} \frac{r}{R}\right) r d r / \int r d r
\end{equation}
where the $\rho_{0}$ = 0.8 is radial flow parameter.

	Particles emitted directly from fireball contain stable and unstable particles. The lifetime of unstable particles is stochastic according to $\exp (-\Gamma \tau)$ exponent in the rest frame of the resonance, and all of them decay into other daughter particles. In the case of two-body decay, the generated daughter particles have momentum in opposite directions in the rest-frame of the resonance, ie.

        \begin{equation}
        \left|\vec{p}_{1}\right| = \left|\vec{p}_{2}\right| = \frac{\sqrt{\left(M^{2}-\left(m_{1}+m_{2}\right)^{2}\right)\left(M^{2}-\left(m_{1}-m_{2}\right)^{2}\right)}}{2 M},
        \label{eq:equation4}
        \end{equation}
    where $M$ is the mass of mother particle, and index 1 and 2 represent two daughter particles.    
    
For $\rm{\overline{{}^4Li}}$ decays into $\rm{\overline{p}}$ and $\rm{\overline{{}^3He}}$, the momentum of the daughter particles is $0.073$ GeV/$c$, which can be derived from Eq.\ref{eq:equation4} where the mass of mother particle ($\rm{\overline{{}^4Li}}$) is 3.751296 GeV (ie. equivalent to the mass corresponding positive particle, $\rm{^4Li}$ ~\cite{nuclear_data}), and the masses of $\overline{p}$ and $\rm{\overline{{}^3He}}$ are 0.93827 GeV and 2.80923 GeV, respectively, in the blast-wave model.
    
The relative abundance of hadrons produced directly is determined by the chemical equilibrium described by a set of parameters including the chemical freeze-out temperature ($T_{ch}$), the baryon chemical potential ($\mu_{B}$) and strangeness chemical potential ($\mu_{S}$)~\cite{song_ratio}:
    \begin{equation}
    n_{i}(T_{ch},\mu_{B},\mu_{S}) = \frac{g_{i}}{2\pi ^{2}}T_{ch}^{3}I(\frac{m_{i}}{T_{ch}},\frac{\mu_{i}}{T_{ch}})
    \end{equation}    
    with $g_{i}$ is the degeneracy factor, and
    \begin{equation}
    \mu_{i} = \mu_{B}B_{i} + \mu_{S}S_{i},
    \end{equation}
    and 
    \begin{align}
    \notag I(\frac{m_{i}}{T_{ch}},\frac{\mu_{i}}{T_{ch}})=
    \int_{0}^{\infty }dx& x^{2}[exp(\sqrt{x^{2}+\frac{m_{i}^{2}}{T_{ch}^{2}}}\\  -&\frac{\mu_{S}S_{i}+\mu_{B}B_{i}}{T_{ch}})\mp 1]^{-1}
    \end{align}
    with upper sign is for bosons and lower sign is for fermions.
    And the probability that a particle belongs to particle type $i$ can be calculated as
    \begin{equation}
    \omega _{i}(T_{ch},\mu_{B},\mu_{S})=\frac{n_{i}(T_{ch},\mu_{B},\mu_{S})}{\sum _{i}n_{i}(T_{ch},\mu_{B},\mu_{S})}.
    \end{equation}

 In the present calculation, the values of the chemical and kinetic freeze-out temperatures [$T_{ch}$ = 0.156 (0.16) GeV 
    	and $T_{k}$ = 0.091 (0.11) GeV for central (peripheral) collisions] as well as  the baryon chemical potential [$\mu_{B}$ = 0.022 (0.019) GeV for central (peripheral) collisions] and strangeness chemical potential [$\mu_{s}$ = 0.0044 (0.0031) GeV for central (peripheral) collisions]  ~\cite{star034909} are selected to be 
    consistent with those from other model calculations  ~\cite{hsong,pbk} as well as 
    the experimentally estimated values ~\cite{alice,c71}.

     From those above equations, we can get the yield ratio $\rm{\overline{{}^4Li}:\overline{{}^4He} = 4.36}$.  Deduced from the ratio of $\rm{\overline{{}^4He}}$ to $\rm{\overline{{}^3He}}$ measured by the STAR Collaboration is $3.2\times10^{-3}$~\cite{naturexue}, we can get  $\rm{\overline{{}^4Li}:\overline{{}^3He} = 0.0148}$. On the other hand, in our final state phase space, $\rm{\overline{{}^4Li}}$ decays into $\rm{\overline{p}}$ and $\rm{\overline{{}^3He}}$ ($ \rm{\overline{{}^4Li}\rightarrow \overline{p} +  \overline{{}^3He} \label{Ham}}$) with the width of 6 MeV ~\cite{hbt87}.

According to the blast-wave model, the momentum information of the final hadrons is obtained. In the STAR experiment, the tracks of particles are reconstructed by the Time Projection Chamber (TPC) ~\cite{resolution} and Heavy Flavor Tracker (HFT) ~\cite{hf2}. According to the momentum resolution of particles in TPC and HFT, we assume a momentum resolution of $1.5\%$ for $\rm{\overline{p}}$ and  $2\%$ for $\rm{\overline{{}^3He}}$ for the phase-space ~\cite{hf1,hf2,resolution}. Based on the resolution, the momentum of particles from the model is smeared. The emission source of high energy heavy ion collisions can be considered spherically symmetric ~\cite{p_lambda}. For Au + Au collisions at $\sqrt{S_{NN}}$ = 200 GeV, the sizes of emission source corresponding to  central collisions and peripheral collisions are different. The radius of emission source for 200GeV Au+Au collision is about 5 - 6 fm according to the STAR experimental results ~\cite{radius1,radius2}. Here we assume the source radius in our case is 5 fm for central collisions. While the peripheral collision has a typical source radius of 3 fm ~\cite{radius1,radius2}.

\begin{figure}[htb]
		\includegraphics[scale=0.46]{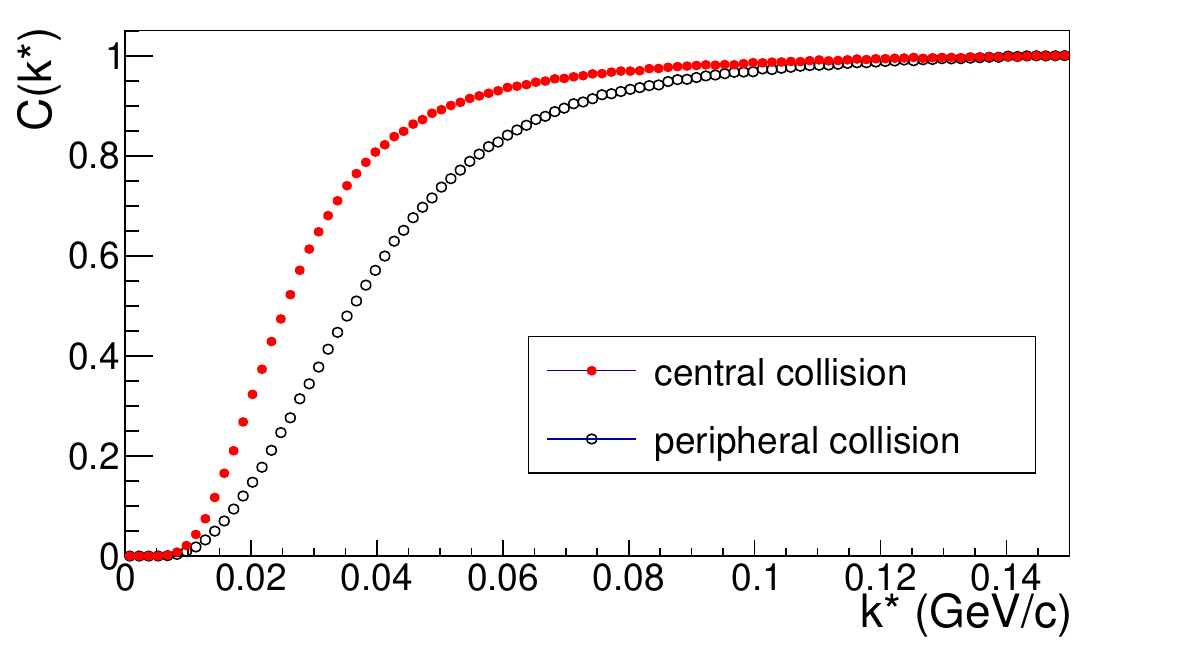}
		\caption{Simulated correlation functions of  $\rm{\overline{p}}$ and $\rm{\overline{{}^3He}}$ in central collisions  ($R$ = 5 fm, open dark circles) and peripheral collisions ($R$ = 3 fm, filled red circles). 
		}
		\label{com_rad}
	\end{figure}

 We assume a spherically symmetric Gaussian distribution for the phase-space and the correlation functions for two different cases are shown in Fig. \ref{com_rad}. There is no decay contribution here, it is only used to discuss the correlation function of $\rm{\overline{p}}$ and $\rm{\overline{{}^3He}}$ which is the background of our measurement.
In the range where the relative momentum between the $\rm{\overline{p}}$ and $\rm{\overline{{}^3He}}$ pairs is small, the correlation function is below 1 due to the repulsive Coulomb interaction between the two particles. One can see that the correlation  becomes weaker as the size of the source increases, which is consistent with the prediction for non-identical particle pairs using Coulomb wave functions only \cite{inter}.

    \section{RESULTS AND DISCUSSION}
    \label{sec:RESULTS}

    In this section, we present the results of the simulated correlation functions for $\rm{\overline{p}}$-$\rm{\overline{{}^3He}}$ with/without $\rm{\overline{{}^4Li}}$ decay. The correlation functions are calculated according to Eq.~\ref{eq:equation13}.  As a useful contrast, the correlation function for $\rm{{proton}}$-$\rm{{}^3{He}}$ derived from the phase space with/without $\rm{{}^4Li}$ is also presented.
	
	\begin{figure}[htb]
		\includegraphics[scale=0.44]{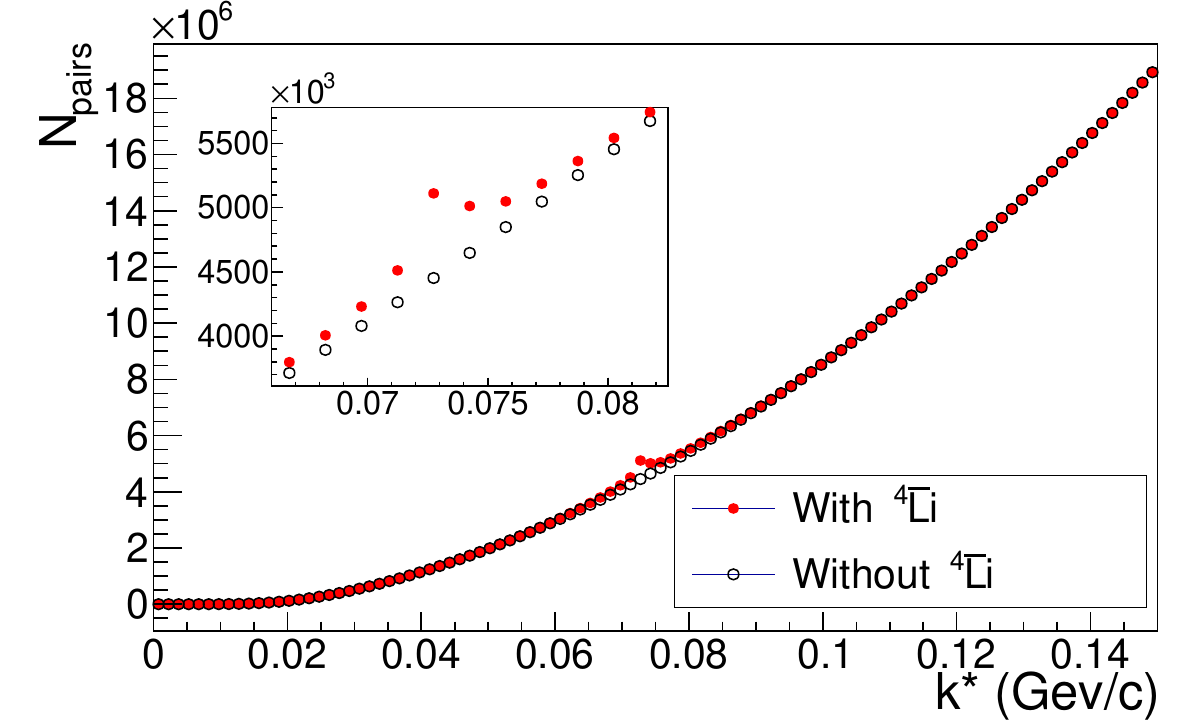}
		\caption{The $k^{*}$ distributions for pairs from same events in the case of central collisions. The events with/without $\rm{\overline{{}^4Li}}$ decay are generated by the blast-wave model. The filled red or open dark circles correspond to the results from phase spaces with or without $\rm{\overline{{}^4Li}}$ decay.}
		\label{Npairs}
	\end{figure}

		\begin{figure}[htb]
			\includegraphics[scale=0.44]{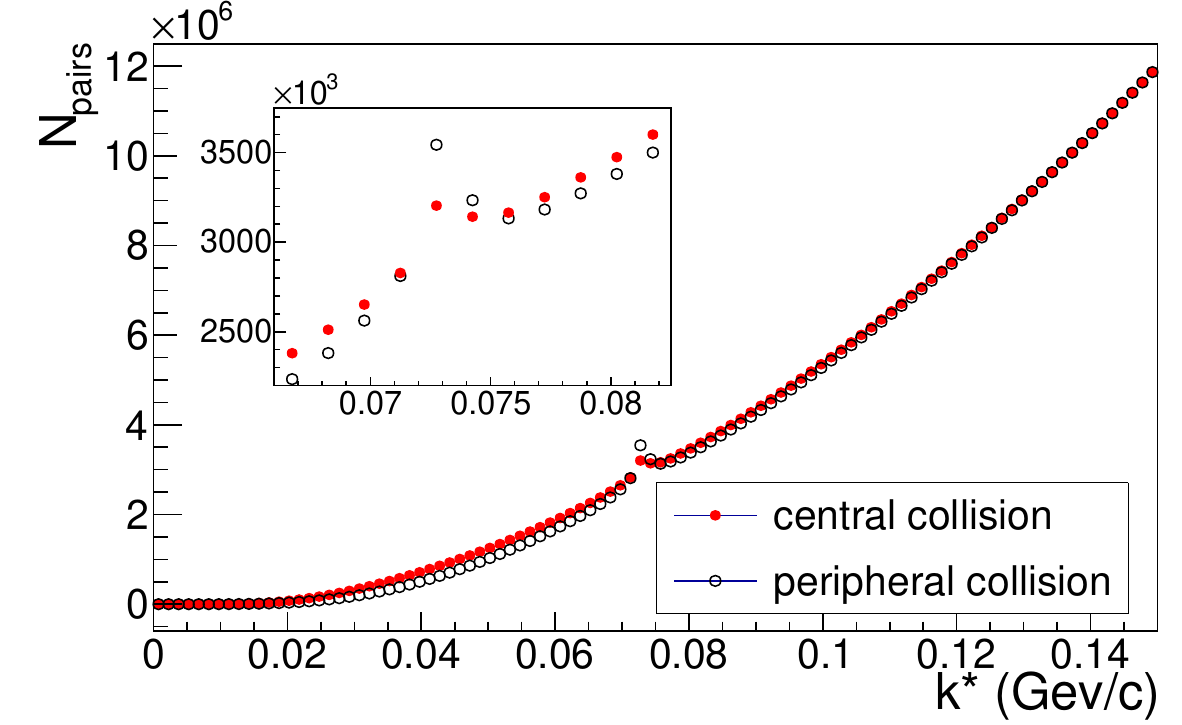}
			\caption{The $k^{*}$ distributions for pairs from same events are shown. The events from central or peripheral collisions are generated by the blast-wave model. The filled red or open dark circles correspond to the central or peripheral collisions.}
			\label{Npairs_p}
		\end{figure}

    \subsection{$k^{*}$ distributions of pairs from the same events}
    
    We generate events using a blast-wave model described in Sec. \ref{subsec:cf} and \ref{subsec:dragon}. To compare the difference between correlation functions with/without $\rm{\overline{{}^4Li}}$ decay, the corresponding phase spaces are produced.
    In one case $\rm{\overline{{}^4Li}}$ is generated in the emission source and decayed, while in the other case $\rm{\overline{{}^4Li}}$ is not generated.
    In both cases we apply the mixed event technique while the weights for the pairs from same events are calculated based on Eq.~\ref{eq:equation133} \ref{eq:equation13}. Fig.~\ref{Npairs} shows the $k^*$ distributions from phase space with/without $\rm{\overline{{}^4Li}}$ decay for central collisions. In the $k^*$ distribution containing $\rm{\overline{{}^4Li}}$ decay, there is a peak at $k^*$ around 0.073 GeV/$c$. The difference will be reflected in the calculated correlation function according to Eq.~\ref{eq:equation13}. Fig.~\ref{Npairs_p} displays a comparison between the central collisions and the peripheral collisions. As expected from the decreased source size, an enhanced peak is observed for peripheral collision. Here the ratio of the multiplicity of the antiproton produced by the central collisions and the peripheral collisions is set as 3.224 according to the STAR data~\cite{star034909}.
     
	        \begin{figure}[htb]
	        		\centering
	        	\includegraphics[scale=0.47]{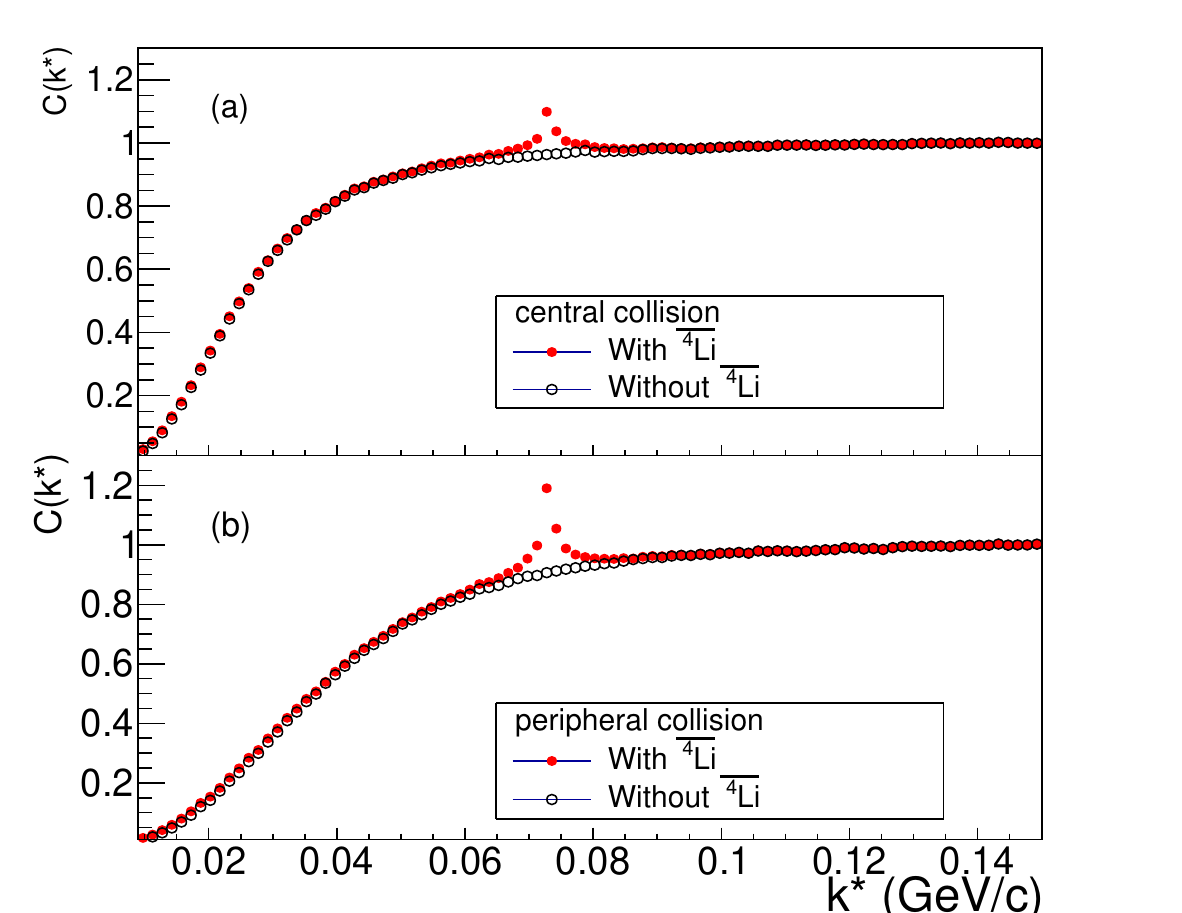}
	        	\caption{The prediction of correlation functions with/without $\rm{\overline{{}^4Li}}$ for central (a) and peripheral (b) Au+Au collisions at $\sqrt{S_{NN}}$ = 200 GeV. The filled red or open dark circles correspond to the correlation function with or without  $\rm{\overline{{}^4Li}}$ decay. 
		A significant peak at $k^*$ around 0.073 GeV/$c$ is shown for the correlation function containing $\rm{\overline{{}^4Li}}$ decay. Here the yield ratio of  $\rm{\overline{{}^4Li}/\overline{{}^4He}}$ is assumed to be 4.36.}
	        	\label{CF_anti}	
	        \end{figure}

   \begin{figure}[htb]
   	\centering
	\includegraphics[scale=0.47]{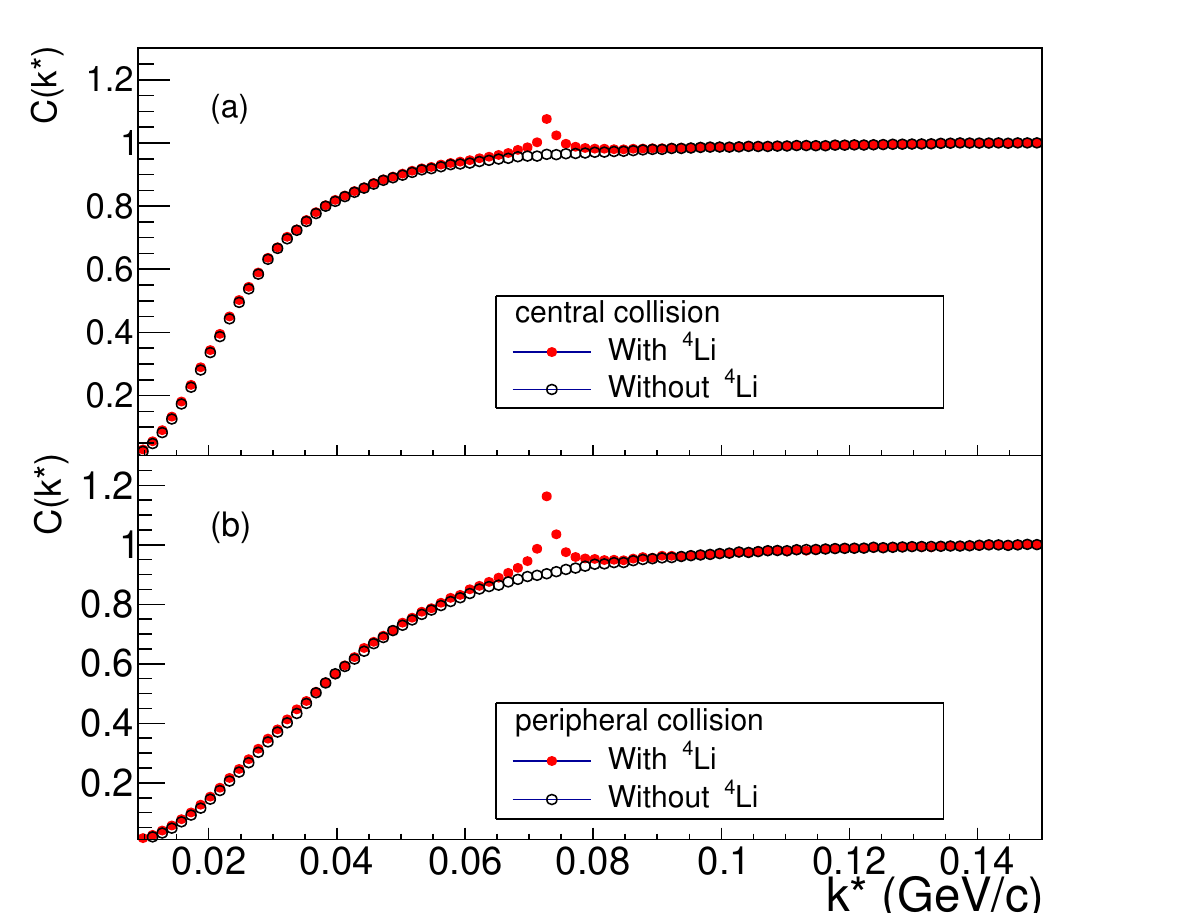}
	\caption{ 
	Same as Fig.~\ref{CF_anti}, but for the correlation functions of $\rm{p}$-$\rm{{}^3{He}}$ with/without $\rm{{}^4{Li}}$.  Here the ratio of $\rm{\overline{p}/p}$ in  the central (peripheral) collisions is set as 0.77 (0.8) according to the STAR data~\cite{star034909}.
	}
	\label{CF_Li4}
\end{figure}

\begin{figure}[htb]
	\includegraphics[scale=0.47]{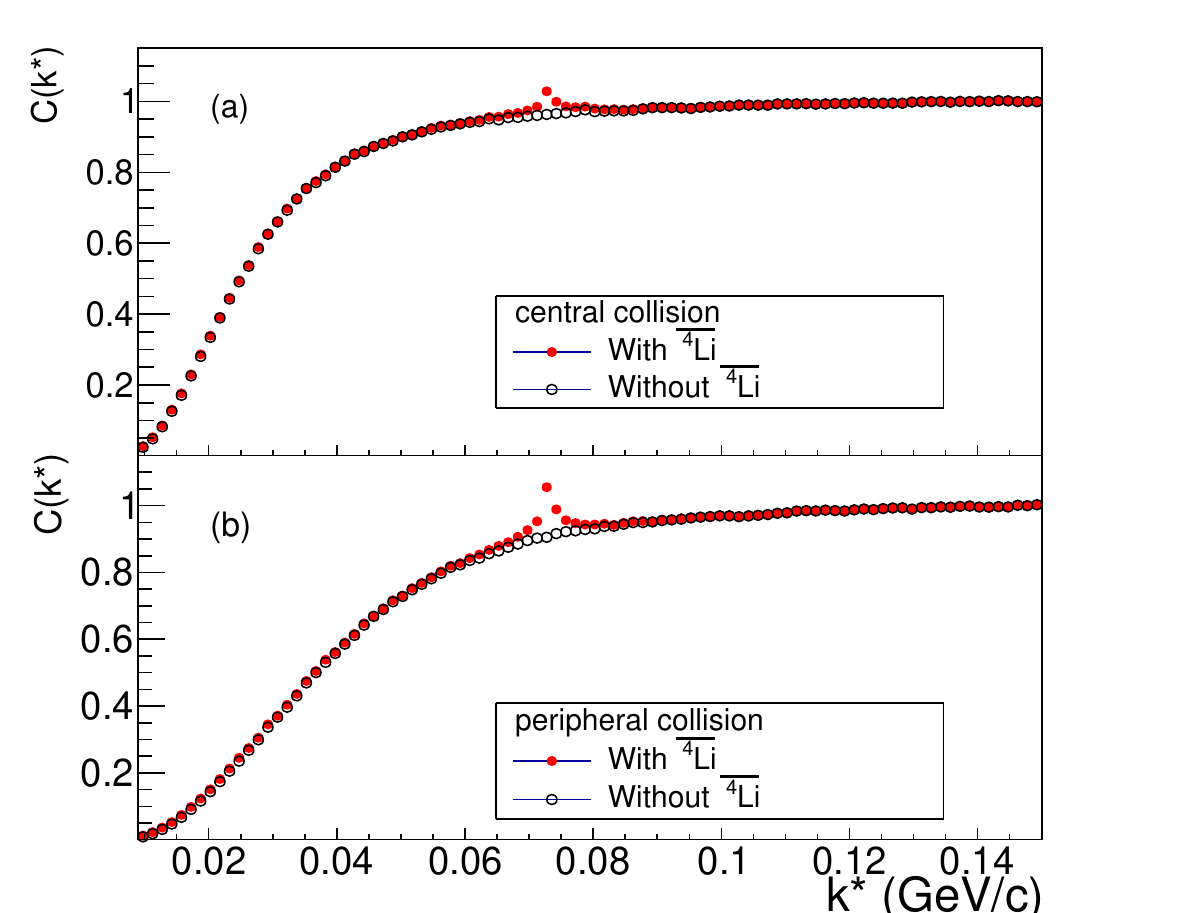}
	\centering
	\caption{ Same as Fig.~\ref{CF_anti}, but assuming the yield ratio of $\rm{\overline{{}^4Li}} / \rm{\overline{{}^4He}}$ is 1:1.}
	\label{CF_low}
\end{figure}
	
	\subsection{Correlation function of $\rm{\overline{p}}$-$\rm{\overline{{}^3He}}$}
	
	Fig.~\ref{CF_anti} shows the prediction of correlation functions with/without $\rm{\overline{{}^4Li}}$ for central (peripheral) Au+Au collisions at $\sqrt{S_{NN}}$ = 200 GeV. The effect of the Coulomb interaction between $\rm{\overline{p}}$ and $\rm{\overline{{}^3He}}$ dominates the correlation functions.
		 The size of our emission source is relatively large due to the central collisions, so the short-range strong interaction between $\rm{\overline{p}}$ and $\rm{\overline{{}^3He}}$ has  little effect to our correlation function. 
		 The correlation function containing $\rm{\overline{{}^4Li}}$ decay is shown as filled red circles in Fig.~\ref{CF_anti}. The upper panel shows a significant peak at $k^*$ around 0.073 GeV/$c$ in comparison with the correlation function without $\rm{\overline{{}^4Li}}$ decay in central collisions. The lower panel of Fig.~\ref{CF_anti}  shows that for the peripheral collisions, the position of the peak of the correlation function containing $\rm{\overline{{}^4Li}}$ decay keeps almost the same due to the same decay kinematics. The strength of the peak is actually determined by the ratio of  $\rm{\overline{{}^3He}}$ and $\rm{\overline{{}^4Li}}$ yield  in the same collision system in our phase space. With larger relative yield for $\rm{\overline{{}^4Li}}$, we would expect stronger peak in our correlation function. Therefore, we can principally measure the  $\rm{\overline{{}^4Li}}$ yield by measuring the correlation function of $\rm{\overline{p}}$-$\rm{\overline{{}^3He}}$. 
		
	According to the error of the obtained correlation function, the number of events required for experimental measurement of $\rm{\overline{{}^4Li}}$ can be estimated on an order of magnitude.	
	When the number of counted events is larger, the error of the correlation function will of course  become smaller. Assuming that error reaches one third of the height of the signal peak, it shall be difficult to see the $\rm{\overline{{}^4Li}}$ signal. According to the assumption, about 1 billion 200 GeV Au+Au  events are required for the experiment.	However, due to the effect of detector efficiency, the number of events required for experiments may be underestimated.
	
	As a comparison, we also show the correlation functions of $\rm{proton}$-$\rm{{}^3{He}}$ with/without $\rm{{}^4{Li}}$ in Fig. \ref{CF_Li4}.  Here the ratio of the multiplicity of the antiproton and the proton produced in the central (peripheral) collisions is set as 0.77 (0.8) according to the STAR data~\cite{star034909}. They show a very similar structure as the $\rm{\overline{p}}$-$\rm{\overline{{}^3He}}$ correlation functions.

Finally, due to the effect of coalescence~\cite{li4h4} and detector efficiency, the yield of $\rm{\overline{{}^4Li}}$ in real experiments might be even lower. Thus we discuss the scenario when $\rm{\overline{{}^4Li}}$ yield is lower. Here we adjust the yield ratio of $\rm{\overline{{}^4Li}}$ and $\rm{\overline{{}^4He}}$ to 1:1, then we can obtain the ratio $\rm{\overline{{}^4Li}:\overline{{}^3He} = 0.0079}$. By using the same method, the correlation function is obtained as shown in Fig.~\ref{CF_low}, from which we can see that there is still a tiny signal which  is weaker in comparison with the one with a higher $\rm{\overline{{}^4Li}}$ yield. This shows that the signal decreases with the decreases of production rate of $\rm{\overline{{}^4Li}}$, but it is still observable. In this scenario, the number of events required for experimentally observing $\rm{\overline{{}^4Li}}$ is about 5 billion.

    \section{SUMMARY}
    \label{sec:SUMMARY}

We use the blast-wave model and Lednick\'y-Lyuboshitz analytical model to obtain a prediction of the correlation function of  $\rm{\overline{p}}$-$\rm{\overline{{}^3He}}$ with/without $\rm{\overline{{}^4Li}}$ decay in Au+Au collisions at $\sqrt{S_{NN}}$ = 200 GeV. The repulsive Coulomb interaction dominates the $\rm{\overline{p}}$-$\rm{\overline{{}^3He}}$ correlation function at lower relative momentum for central collisions. The correlation function with $\rm{\overline{{}^4Li}}$ decay is found to exhibit a peak at $k^* \approx$ 0.073 GeV/$c$. And  the event number required for experimentally detection of $\rm{\overline{{}^4Li}}$ is estimated. The present study sheds light on an experimental search for $\rm{\overline{{}^4Li}}$ in relativistic heavy ion collisions.

\begin{acknowledgements}
	We thank Dr. Lednick\'y for the useful discussion on the calculation of the correlation function. This work is partially supported by the National Natural Science Foundation of China under Contract Nos. 11890714, 11421505,
	11875066, 11925502 and 11961141003, National Key
	R\&D Program of China under Grant No. 2016YFE0100900 and
	2018YFE0104600, the Key Research Program of Frontier Sciences
	of the CAS under Grant No. QYZDJ-SSW-SLH002, and the Key Re-
	search Program of the CAS under Grant NO. XDPB09.

\end{acknowledgements}
~\\

\renewcommand\thesection{APPENDIX~\Alph{section}}
\renewcommand\theequation{\Alph{section}.\arabic{equation}}

\end{CJK*}
\end{document}